# Interfacial Thermal Conductance across Room-Temperature Bonded GaN-Diamond Interfaces for GaN-on-Diamond Devices

Zhe Cheng,[1,a)] Fengwen Mu,[2,3,a),*] Luke Yates,[1] Tadatomo Suga,[2] Samuel Graham[1,4,*]

[1] George W. Woodruff School of Mechanical Engineering, Georgia Institute of Technology, Atlanta, Georgia 30332, USA

[2] Collaborative Research Center, Meisei University, Hino-shi, Tokyo 191-8506, Japan

[3] Kagami Memorial Research Institute for Materials Science and Technology, Waseda University, Shinjuku, Tokyo 169-0051, Japan

[4] School of Materials Science and Engineering, Georgia Institute of Technology, Atlanta, Georgia 30332, USA

[a)] These authors contribute equally

[*] Corresponding authors: mufengwen123@gmail.com; sgraham@gatech.edu

**Abstract**

The wide bandgap, high-breakdown electric field, and high carrier mobility makes GaN an ideal material for high-power and high-frequency electronics applications such as wireless communication and radar systems. However, the performance and reliability of GaN-based high electron mobility transistors (HEMTs) are limited by the high channel temperature induced by Joule-heating in the device channel. High thermal conductivity substrates (e.g., diamond) integrated with GaN can improve the extraction of heat from GaN-based HEMTs and lower the device operating temperature. However, heterogeneous integration of GaN with diamond substrates is not trivial and presents technical challenges to maximize the heat dissipation potential brought by the diamond substrate. In this work, two modified room-temperature surface-activated bonding (SAB) techniques are used to bond GaN and single crystal diamond with different interlayer thicknesses. Time-domain thermoreflectance (TDTR) is used to measure the thermal properties from room temperature to 480 K. A relatively large thermal boundary conductance (TBC) of the GaN-diamond interfaces with a ~4-nm interlayer (~90 MW/m$^2$-K) was observed and material characterization was performed to link the structure of the interface to the TBC. Device modeling shows that the measured GaN-diamond TBC values obtained from bonding can enable high power GaN devices by taking the full advantage of the high thermal conductivity of single crystal diamond and achieve excellent cooling effect. Furthermore, the room-temperature bonding process in this work do not induce stress problem due to different coefficient of thermal expansion in other high temperature integration processes in previous studies. Our work sheds light on the potential for room-temperature heterogeneous integration of semiconductors with diamond for applications of electronics cooling especially for GaN-on-diamond devices.

## Introduction

With wide bandgap, high-breakdown electric field, and high carrier mobility, GaN has been used for high-power and high-frequency electronics applications such as wireless communication, satellite communication, and radar systems.[1] The maximum output power of GaN-based HEMTs is limited by the high channel temperature induced by localized Joule-heating, which degrades device performance and reliability.[2,3] Diamond has the highest thermal conductivity among natural materials and is of interest for integration with GaN to help dissipate the generated heat from the channel of GaN-based HEMTs.[2,4-6] Current techniques involve two ways to integrate GaN with diamond. One is direct growth of chemical vapor deposited (CVD) diamond on GaN with a transition layer of dielectric material.[7] The nanocrystalline diamond near the nucleation interface has reduced thermal conductivity (tens of W/m-K) which could contribute to an additional thermal resistance of 10 $m^2$K/GW.[8] Combining with the GaN-diamond thermal boundary resistance, these near-hotspot thermal resistances have been shown to have a large impact on impeding the flow of heat from the device channels, especially for high frequency applications in which the thermal penetration depth is small.[9-12] The high growth temperature of the diamond also induces large residual stress in the GaN because of the mismatch of the coefficients of thermal expansion.[13,14] Stresses generated during the direct growth of diamond on GaN have been shown to vary, dependent upon GaN thickness, diamond growth temperature, and the sacrificial carrier wafer[15,16]. Residual stresses greater than 1 GPa at the free surface of the GaN have been reported.[17] The elevated stress conditions in the GaN ultimately limit the total thickness and material quality of the GaN by inducing layer cracking and wafer bow, and impact the electrical performance of the device.[18-21] The aforementioned effects of elevated stress cause a significant reliability concern when considering the function and lifetime of a GaN device. Another method is high temperature bonding of GaN with diamond.[22,23] The GaN device is firstly grown on silicon substrates. Then the GaN device is transferred and bonded with a CVD diamond with an adhesion layer. The transfer and bonding processes are performed at elevated temperatures exceeding 700 °C.[23] The adhesion layer increases the thermal resistance of GaN-diamond interface, which offsets the effect of the high thermal conductivity of diamond substrates. The stress due to the different coefficients of thermal expansion results in wafer bow and warp, even fractures.[23] Even though some attempts have been made to bond GaN with diamond at lower temperatures[4,24], additional techniques to integrate GaN with diamond substrates are in demand to be developed to take full advantage of the high thermal conductivity of diamond without inducing additional stress resulting from high temperature processes.

In this work, we use two modified SAB techniques (one with Si nanolayer sputtering deposition and the other with Si-containing Ar ion beam) to bond GaN with diamond substrates with different interlayers at room temperature. TDTR is used to measure the thermal properties. Materials characterization such as high-resolution scanning transmission electron microscopy (HR-STEM) and electron energy loss spectroscopy (EELS) are used to study the interface structure and chemistry to help elucidate the measured thermal properties. An analytical modeling for devices is performed to estimate the device cooling performance of these bonded interfaces.

## Results and Discussion

In this work, GaN was bonded to diamond using a Si interlayer to aid in the chemical adhesion at the interface. The first sample, Samp1, is comprised of a thin layer of GaN (~700 nm) bonded on a commercial single crystal diamond substrate (grown by CVD and purchased from EDP

Corporation) with ~10-nm-thick Si interlayer. The Si interlayer can lower TBC, so a different bonding method is applied to Samp2 which has a ~1.88-μm-thick GaN bonded onto a commercial single crystal diamond substrate grown by a high-pressure high-temperature (HPHT) method and purchased from Sumitomo Electric Industries, Ltd. In this sample, the Si-containing Ar ion beam is employed to introduce a ~2-nm-thick interlayer to enhance the interfacial chemical interaction between GaN and diamond. More details about the bonding process can be found in Experimental Section and references[25,26].

To create the sample for measuring the TBC between the GaN and diamond by TDTR as shown in Figure 1(a), a layer of Al was deposited on the sample surface as a TDTR transducer. With this sample structure, the thermal conductivity of the single crystal diamond substrates were measured first on the area without GaN. Then TDTR measurements were performed on the area with the GaN layer to measure the GaN-diamond TBC. The measured thermal conductivity of the diamond substrates was used as a known parameter in the TDTR data fitting to extract the TBC when measuring over the GaN layer. Overall, there are three unknown parameters: Al-GaN TBC, GaN thermal conductivity, and GaN-diamond TBC. As shown in Figure 1 (b), the TDTR sensitivity of these three unknown parameters are large which is good for accurate thermal measurements. TDTR is a pump-probe technique which can measure thermal properties of both nanostructured and bulk materials.[27,28] A modulated pump laser beam heats the sample surface while a delayed probe beam detects the surface temperature variation via thermoreflectance.[11] The signal which is picked up by a photodetector and a lock-in amplifier is fitted with an analytical heat transfer solution to infer unknown parameters. More details about TDTR measurements on similar sample structures can be found in the literature.[10,26,28,29] An example of the agreement between experimental data (Exp) and the analytical heat transfer solution (Theo) in TDTR data fitting is shown in Figure 1 (c).

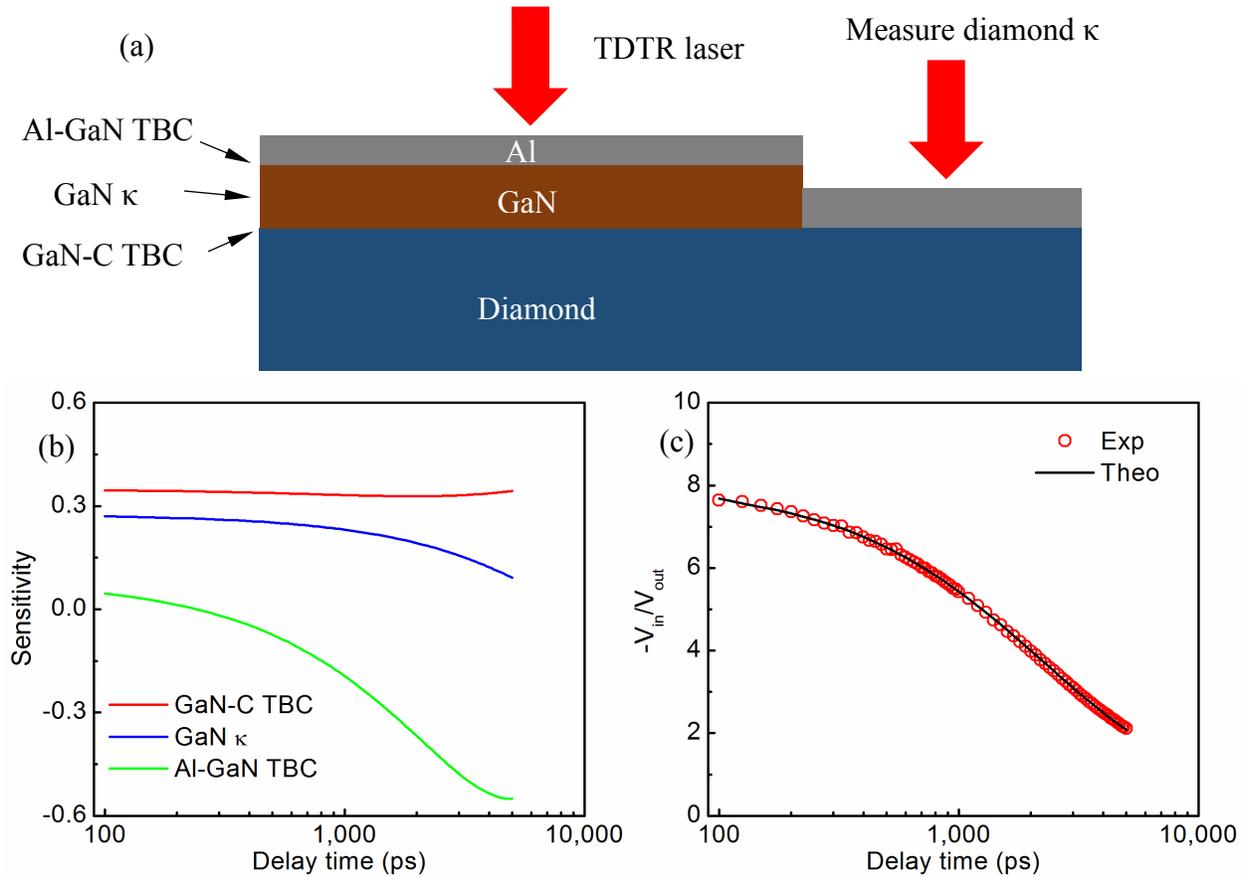

Figure 1. (a) TDTR measurements on the diamond and bonded GaN-diamond samples. (b) TDTR sensitivity of the three unknown parameters of Samp2. (c) TDTR data fitting of Samp2 with modulation frequency of 2.2 MHz at room temperature.

Figure 2(a-b) show the temperature dependence of the measured thermal conductivity of the diamond and the GaN layer and are compared with literature values. Our measured diamond and GaN thermal conductivity match with literature values.[30-34] The light yellow color of HPHT sample shows relatively high concentration of impurities while the CVD diamond sample used in this work is transparent. The CVD diamond sample has a higher thermal conductivity than the

HPHT sample as expected and matches well with literature values of high-purity diamonds. [30,31] The measured thermal conductivity of the GaN layer (~1.88 μm) in Samp2 is close to experimentally measured bulk values and lower than density-function-theory calculated (DFT) values because of impurities in these samples. The slightly larger thermal conductivity difference between experimental values and DFT values at high temperatures is due to higher order of phonon scatterings such as four phonon scattering.[32] Phonon-phonon scattering dominates in thermal transport at high temperatures. Because of the large phonon bandgap of GaN (the optical phonons have much larger energy than the acoustic phonons), the three-phonon scattering process among acoustic and optical phonons are limited (energy conservation during phonon scattering process). Four-phonon scattering process which is not included in the DFT calculation in the reference[33] becomes relatively important, leading to an overestimated thermal conductivity. The thermal conductivity of the GaN thin film (~700 nm) in Samp1 does not have good TDTR sensitivity so literature values are used in the TDTR data fitting.[26,35]

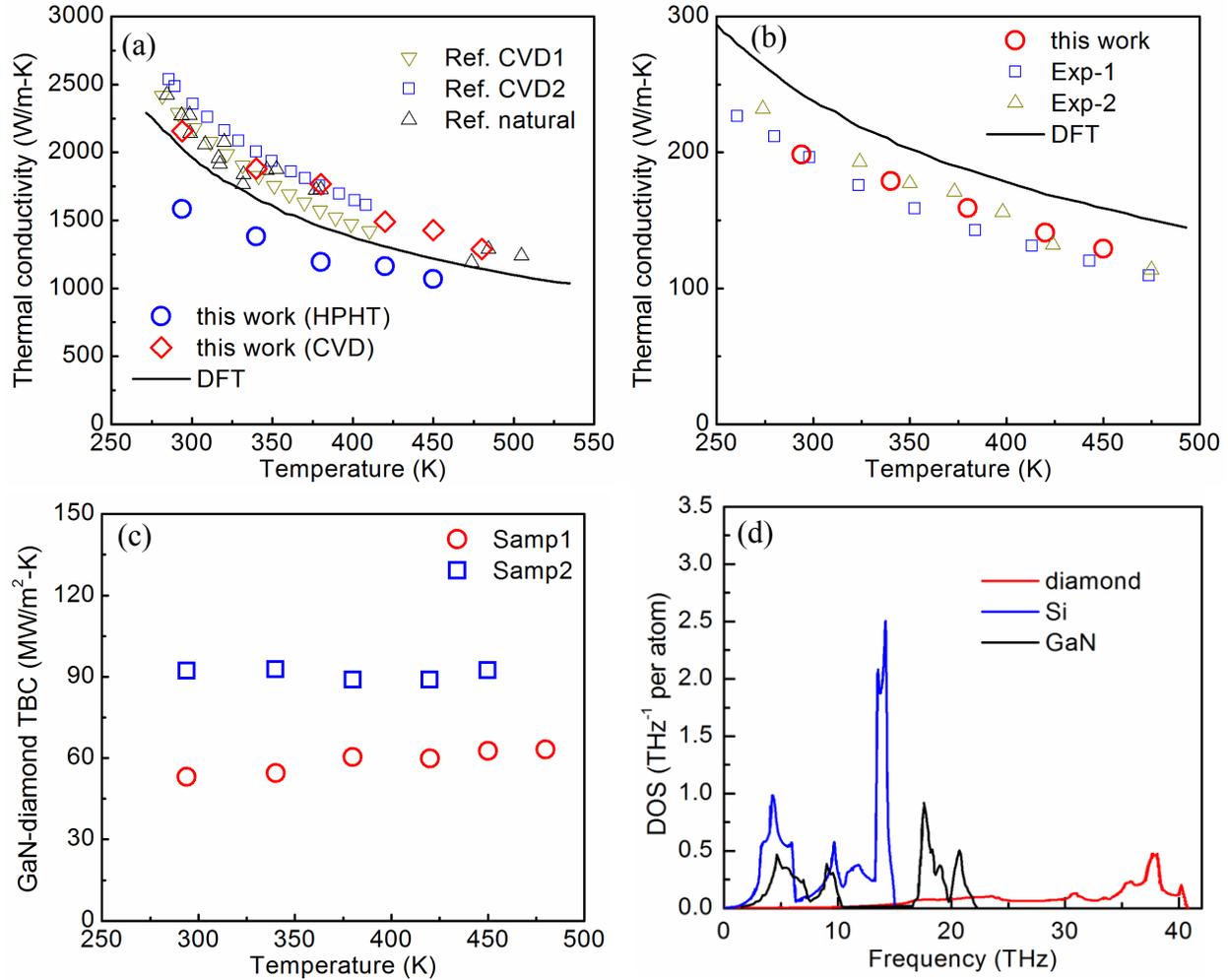

Figure 2. (a) temperature dependence of the measured thermal conductivity of two diamond substrates: Samp1 (CVD) and Samp2 (HPHT). The DFT values and the measured thermal conductivity of Ref. CVD1, CVD2, and natural are from literature.[30,31] (b) temperature dependence of the measured thermal conductivity of GaN layer. "DFT", "Exp-1", and "Exp-2" are DFT-calculated thermal conductivity and two experimentally measured thermal conductivity of bulk GaN from literature.[32-34] (c) temperature dependence of the measured TBC of bonded GaN-diamond interfaces. (d) phonon density of state of GaN, Si, and diamond.[10,36]

The temperature dependence of the measured TBC of room-temperature bonded GaN-diamond interfaces are shown in Figure 2(c). Samp2 has a much higher TBC (92 MW/m$^2$-K) than Samp1 (53 MW/m$^2$-K) at room temperature because of the thinner interlayer. For the data fitting of the TDTR measurements in this work, the Si interlayer is very thin, so its thermal resistance is added to the total thermal resistance of the interface. The measured TBC is the effective TBC of the GaN-Si-diamond architecture at the interface (GaN-Si interface + Si layer + Si-diamond interface). The measured TBC of both samples show weak temperature dependence in the measured temperature range (295 K to 480 K). The Debye temperatures of GaN, Si, and diamond are higher than 480 K.[10,37] For perfect GaN-Si-diamond interfaces, effective TBC should increase with temperature in the range of 295 K to 480 K because an increasing number of phonons are involved in the thermal transport across the GaN-diamond interfaces as temperature increases. However, for the bonded GaN-Si-diamond interfaces in this work, the disorder at the interface and Ar atoms trapped at the interface increases phonon scattering, which possibly explains the weak temperature dependence of the measured TBC. More details about the interfacial structure-thermal property relationship will be discussed later. Figure 2(d) shows the phonon density of states (DOS) of GaN, Si, and diamond.[10,36] Generally speaking, large phonon DOS overlap leads to large TBC. A thin interlayer with a max phonon frequency between two adjacent materials has been reported to cause an increase in TBC.[38,39] As shown in Figure 2(d), the max phonon frequency of Si is lower than those of both GaN and diamond. The Si interlayer would reduce the TBC of GaN-diamond interfaces. Therefore, the TBC of bonded GaN-diamond interfaces still have the potential to be improved by using other interfacial layers such as SiC, AlN, or SiNx, even though our measured TBC for Samp2 is already among the high TBC for GaN-diamond interfaces.[6,40]

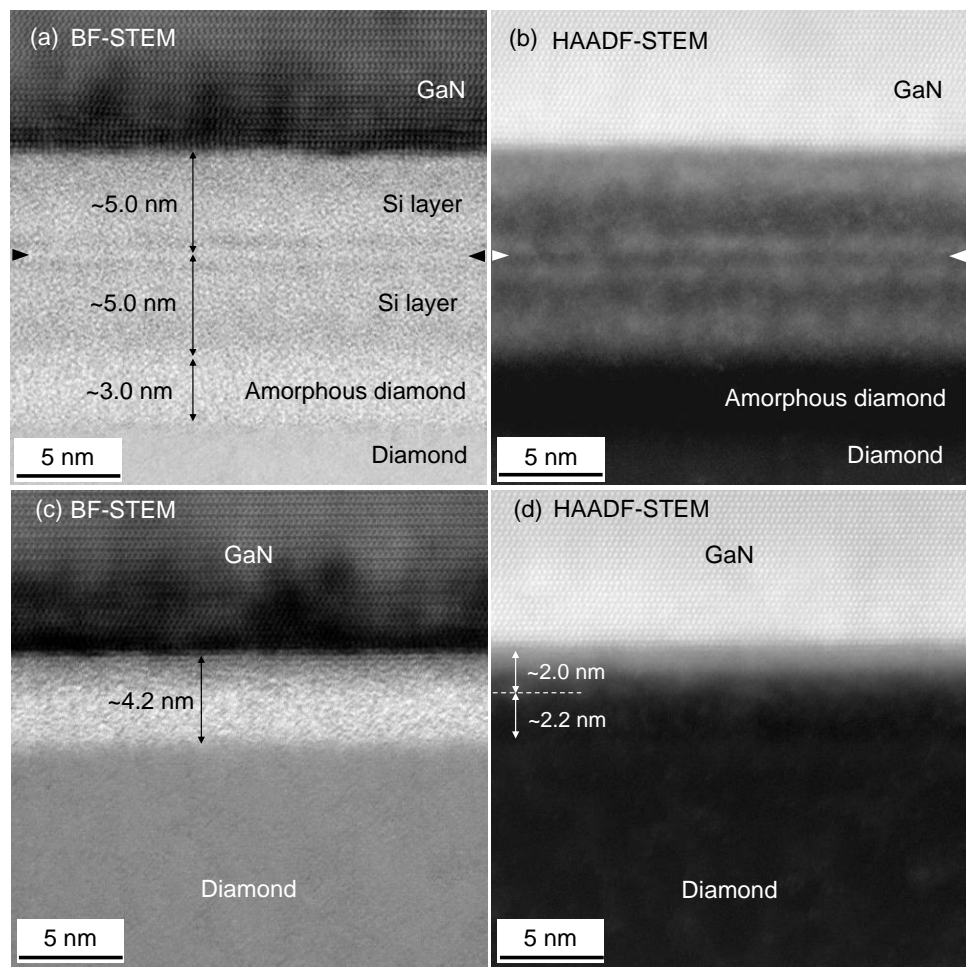

Figure 3. (a-b) Cross-section bright-field (BF) and high-angle annular dark-field (HAADF) HR-STEM images of GaN-diamond interfaces of Samp1. (c-d) Cross-section bright-field (BF) and high-angle annular dark-field (HAADF) HR-STEM images of GaN-diamond interfaces of Samp2.

To help elucidate the measured TBC and its relationship to the sample architecture, high-resolution scanning transmission electron microscopy (HR-STEM) and electron energy loss spectroscopy (EELS) are used to study the structure of the GaN-diamond interfaces. As shown in Figure 3(a-b), the GaN-diamond interfaces of Samp1 are composed as two layers of amorphous deposited silicon and one amorphous diamond layer. Similar to a previous study in the literature[25], no amorphous

GaN is observed. The bonding interface is the Si-Si interface, marked by two triangles in Figure 3(a-b). The thicknesses of the Si interlayer and the amorphous diamond induced by the surface activation with an Ar ion beam are ~10 nm and ~3.0 nm, respectively. Figure 3(c-d) shows the BF and HAADF HR-STEM images of the GaN-diamond interface of Samp2. Only a ~4-nm-thick overall amorphous layer is observed. The bonded interface is not sharp and the thickness of amorphous diamond is approximately 2 nm. The 2-nm-thick amorphous layer between GaN and diamond is supposed to be amorphous silicon deposited by the Si-containing Ar ion beam during surface activation. TEM images of the interfaces with a low magnification are show in Figure S2, showing that both interfaces are very uniform.

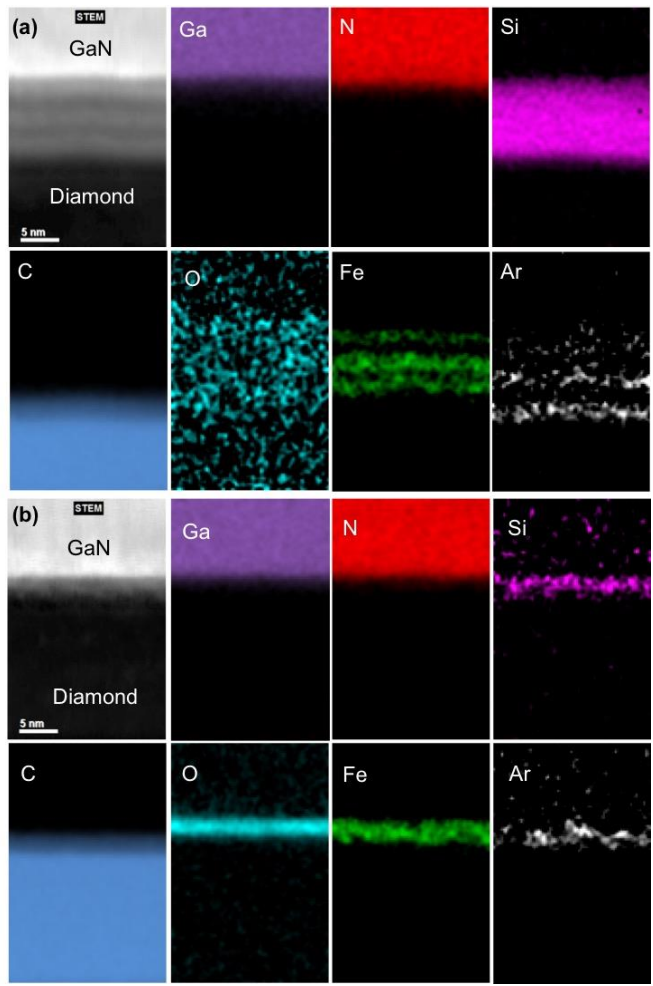

Figure 4. STEM images of the GaN-diamond interface of Samp1 (a) and Samp2 (b), followed by their high-resolution EELS mappings: Ga map in purple, N map in red, Si map in pink, C map in blue, O map in Cyan, Fe map in green, and Ar map in white.

Additionally, high-resolution EELS analysis is used to study the chemical composition at the interfaces. As shown in Figure 4(a), the EELS element mapping of the interface confirms the interlayer thicknesses in the TEM images. Si atoms are implanted into diamond and GaN, and Ar atoms are also implanted into diamond. Some ion elements also show up at the interfaces, originating from the ion beam source which is made of stainless steel. The three-layer distribution of Fe corresponds to the activated GaN surface, the activated diamond surface, and the bonding interface. The bonding interface has the highest atomic composition of Fe (~6%). Please note these Fe contamination could be removed after further improvement of the bonding environment. The O maps in Figure 4(a) is oxygen-contaminated after sample preparation, which is supposed to be no oxygen at the interface. Different from the interface bonded by modified SAB with sputtering-deposited Si nanolayer, Si-containing Ar ion beam does not cause much implantation of Si at the interface. As shown in Figure 4(b), the EELS element mapping of the interface of Samp2 indicates that the interface layer is composed of Si, Ar, O, and Fe. No implantation of Si is observed near the interface. The disorder at the interface confirms our discussion above about the weak temperature dependence of GaN-diamond TBC. Due to the complicated nature of interfacial thermal transport, it is still unclear that how these imperfections such as Ar, O, and Fe defects, and amorphous Si, affect the TBC of the bonded interfaces. Further processing refinements are necessary to change or control the distribution of defects, impurities, and the amorphous layers to further improve TBC and elucidate their effects.

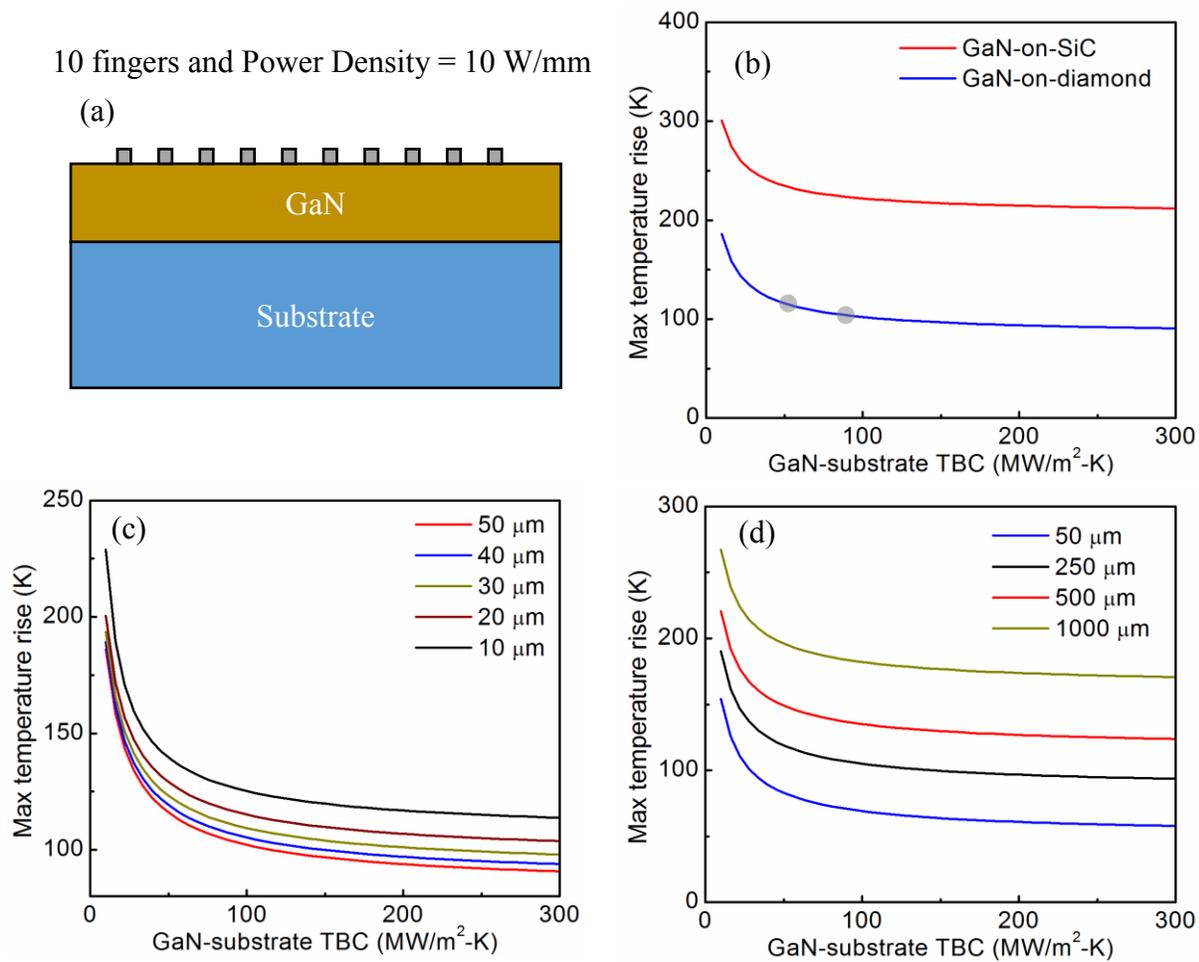

Figure 5. (a) A 800-nm GaN device with 10 fingers seated on a substrate is modeled with a power density of 10 W/mm.[41] (b) the max temperature of the device is calculated with different GaN-substrate TBC and substrates. (c) the effect of gate-gate spacing (10-50 μm) on the max temperature with different GaN-diamond TBC. (d) the effect of gate width (50-1000 μm) on the max temperature with different GaN-diamond TBC.

To estimate the potential effect of room-temperature bonded GaN-Diamond interfaces on the thermal response of GaN HEMTs, a 800-nm GaN device with 10 fingers seated on different substrates is modeled under fully open channel condition with a power density of 10 W/mm.[41] The thermal conductivity of 800 nm layer GaN is 150 W/m-K.[26,35] Figure 5(b) shows the max temperature of the device with a heating source width of 4 µm, a heating source length of 500 µm, and gate-gate spacing of 50 µm. The thermal conductivity of SiC and diamond used in the modeling are 380 W/m-K and 2000 W/m-K, respectively.[32,42] The max temperature of GaN devices on a diamond substrate is much lower than that on a SiC substrate, indicating the advantage of using diamond substrates. Max temperature decreases sharply with increasing GaN-substrate TBC when GaN-substrate TBC is small but saturates for large GaN-substrate TBC. Figure 5(c) shows the effect of gate-gate spacing on the max temperature with different GaN-diamond TBC. The power density, heating source width, gate width of the device are 10 W/mm, 4 µm, and 500 µm, respectively. The gate-gate spacing ranges from 10 µm to 50 µm. The max temperature of devices increases with decreasing gate-gate spacing. When the GaN-diamond TBC is small (<50 MW/m$^2$-K), the max temperature increases more sharply with decreasing GaN-diamond TBC for small gate-gate spacing. Increasing GaN-diamond TBC is very important for devices with reduced gate-gate spacing. Figure 5(d) shows the effect of gate width on the max temperature with different GaN-diamond TBC. The power density, heating source width, gate-gate spacing of the device are 10 W/mm, 4 µm, and 20 µm, respectively. The gate width ranges from 50 µm to 1000 µm. The max temperature increases with gate width while keeping power density constant. The GaN-substrate TBC shows similar trend with Figure 5(b). For all cases modelled in Figure 5(b-d), GaN-diamond TBC is the key to minimize the max temperature especially for small GaN-diamond TBC values. Additionally, the measured TBC of the GaN-diamond interfaces in this work are about 50 and 90 MW/m$^2$-K. The max temperatures of GaN-on-diamond devices with these TBC values are shown as solid gray dots in Figure 5(b). The max temperature does not decrease significantly with TBC if the TBC is larger than 50 MW/m$^2$-K. The max temperature of GaN devices bonded on diamond substrates is much lower than that on SiC substrates, showing the great potential of cooling GaN devices bonded with single crystal diamond substrates. To compare the cooling performance of GaN devices with state-of-the-art GaN-diamond/SiC/Si TBC, we model the max temperature of 800 nm GaN devices with heating source width of 4 μm, gate-gate spacing of 20 μm, and gate width of 500 μm. Even if we use the highest GaN-SiC TBC (230 MW/m$^2$-K) and GaN-Si TBC (143 MW/m$^2$-K) in the literature,[26,43] the power density of GaN-on-diamond can reach 20.3 W/mm, ~2.5 times as that of GaN-on-SiC (8.2 W/mm), and ~5.4 times as that of GaN-on-Si (3.78 W/mm) with a max temperature of 250 °C of the devices.

Table 1. Summary of GaN-diamond TBC in the literature and this work.

| | Method | Conditions | Interlayer | Method | TBC (MW/m$^2$-K) |
|---|---|---|---|---|---|
| Ref.[44] | CVD growth of diamond on GaN | | ~25 nm dielectric | Raman | ~37 |
| Ref.[44] | CVD growth of diamond on GaN | | ~50 nm dielectric | Raman | ~28 |
| Ref.[45] | CVD growth of diamond on GaN | >600 °C | ~50 nm dielectric | Raman | 56 |
| Ref.[46] | CVD growth of diamond on GaN | | ~30 nm SiNx | TDTR | ~35 |
| Ref.[47] | CVD growth of diamond on GaN | | 28 nm SiNx | TTR | 83 |
| Ref.[22] | High temperature bonding | >700 °C | Adhesion layer | TDTR | 21-28 |
| Ref.[48] | High temperature bonding | >700 °C | 22 nm SiNx | TDTR | ~58 |
| Ref.[6] | CVD growth of diamond on GaN | | ~ 5 nm SiNx | TDTR | ~100 |
| Ref.[40] | CVD growth of diamond on GaN | | ~ 5 nm SiNx | TTR | ~150 |
| This work | SAB bonding | Room Temp. | ~ 10 nm Si | TDTR | 53 |
| This work | SAB bonding | Room Temp. | ~ 2 nm Si | TDTR | 92 |

Table 1 summarizes the GaN-diamond TBC measured in the literature and this work. "TTR" is transient thermoreflectance. Our measured GaN-diamond TBC for Samp2 is among the high TBC values reported ever. An advantage of the bonding technique in this work is the room-temperature processing temperature so no additional stress remains due to different coefficient of thermal expansion after bonding.[23] It is notable that the diamond used in this work is single crystal diamond with ultra-high thermal conductivity. However, for CVD growth of diamond on GaN with a dielectric layer, the nanoscrystalline diamond near the GaN-diamond interfaces has significantly reduced thermal conductivity (tens of W/m-K).[9-11,49] Moreover, the thermal conductivity of nanoscrystalline CVD diamond is anisotropic and nonhomogeneous which offsets the high thermal conductivity of diamond for cooling GaN devices. The single crystal diamond substrates used in this work do not have these disadvantages and will pave the way for thermal management of GaN-on-diamond devices.

## Conclusions

This work reported the heterogeneously integration of GaN with single crystalline diamond substrates with two modified room-temperature surface-activated bonding techniques for thermal management of GaN-on-diamond applications. The measured TBC of the bonded GaN-diamond interfaces is among the high values reported in the literatures and is affected by the thickness of the interlayer. Due to the disorder and defects at the interfaces, a weak temperature dependence of GaN-diamond TBC was observed. HR-STEM and EELS results show the presence of interfacial amorphous layers and their compositions at the bonded interfaces. The thermal conductivity of two single crystal diamond substrates and the GaN film were also measured and matches reasonably with literature values. Device modeling shows a relatively large GaN-diamond TBC value (>50 MW/m$^2$-K) achieved by surface activated bonding for GaN-on-diamond devices could enable to take full advantage of the high thermal conductivity of single crystalline diamond. This work paves the way for room-temperature heterogeneous integration of GaN with diamond and will impact applications such as electronics cooling especially for GaN-on-diamond devices.

## Experimental Sections

### Sample Preparation

Two single crystalline diamond (CVD diamond and HPHT diamond with size of 10 mm×10 mm and 3 mm×3 mm) are bonded to templated GaN films at room temperature. The GaN films are Ga-face ~2-μm GaN layer grown on sapphire substrates (~430 μm thick). The root-mean-square (RMS) surface roughness of the GaN and diamond surface is ~0.4 nm and ~0.3 nm, respectively. A modified SAB method with a sputtering-deposited Si nano-layer is used to bond the CVD diamond (Samp1) and a modified SAB method with Si-containing Ar ion beam is used to bond the HPHT diamond (Samp2). The detailed bonding process are similar to literature[25,50]. After bonding, the sapphire substrate was removed by a laser lift-off process. The GaN layer was polished to be thinner to obtain good TDTR sensitivity of the buried GaN-diamond interface. The GaN layer of Samp1 is thinned to ~700 nm, while the GaN layer of Samp2 is thinned to ~1.8 μm. After that, a ~70 nm Al layer was deposited on the samples by sputtering as TDTR transducer.

### TDTR Measurements

The measured thermal conductivity of the diamond substrates are used as input in the data fitting of TDTR measurements on GaN-diamond interfaces. A 10X objective (pump radius: 9.7 μm; probe radius: 5.8 μm) is used with a modulation frequency of 2.2 MHz (Samp2) or 3.6 MHz

(Samp1). The GaN layer of Samp2 is thicker than that of Samp1 so we need large thermal penetration depth to penetrate through the GaN layer to obtain large TDTR sensitivity of the GaN-diamond TBC. Therefore, lower modulation frequency (2.2 MHz) is used for Samp2 to get larger thermal penetration depth. The heat capacity of GaN is from literature.[32] Because of the thin thickness of the GaN layer of Samp1, the GaN thermal conductivity is not sensitive so we fixed the GaN thermal conductivity in the data fitting using literature values of similar GaN layer bonded on SiC.[26] The thermal conductivity and heat capacity of the Al transducer is from literature with similar Al films.[26] The picosecond acoustic technique is used to measure the local Al and GaN thicknesses.[26,51]

**Materials Characterization**

Cross-section TEM samples were prepared with a FEI Helios dual beam focused ion beam (FIB) system. The interface structures were characterized by a HR-STEM (Probe-corrected FEI Titan) and the interface composition was measured by EELS (Gatan Enfinium) with a step size of 0.2 nm. The observation in this study is along <11-20> axis of GaN.

**Supplementary Materials**

The supplementary materials include the TEM images of the bonded interface with low magnification.

**Acknowledgements**

Z. C., L. Y., and S. G. would like to acknowledge the financial support from U.S. ONR MURI Grant No. N00014-18-1-2429. F.M., and T.S. would like to acknowledge the financial support from R&D for Expansion of Radio Wave Resources, organized by the Ministry of Internal Affairs and Communications, Japan and JSPS KAKENHI Grant Number 19K15298.

**Completing Financial Interest**

The authors claim no completing financial interest.